# Unexplored Aspect of Velocity of Light

**Abhijit Biswas, Krishnan RS Mani** [*]

Department of Physics, Godopy Center for Scientific Research, Calcutta 700 008, India

**Abstract**

In the post-Maxwellian era, sensing that the tide of discoveries in electromagnetism indicated the decline of the mechanical view, Einstein replaced Newton's three absolutes**:** space, time and mass – by a single one**:** the velocity of light. The magnitude of the velocity of light was first determined and proven to be finite independently by Ole Römer and Bradley in the eighteenth century. In the nineteenth century, Fizeau carried out the first successful measurement of the speed of light using an earthbound apparatus. Thereafter, many earthbound experiments were conducted for its determination till 1983, when its value was frozen at a fixed value after it was determined up to an accuracy level of a fraction of a meter per second. Einstein considered the speed of light derived mostly from terrestrial experiments, to be the limiting speed of all natural phenomena. Einstein stated in connection with his general relativity theory that light rays could curve only when the velocity of propagation of light varies with position. Experiments have been conducted to prove the phenomena of light deflection to higher and higher accuracy levels. But till date, no experiment has been conducted to determine the speed of light at locations closer to the sun even after several decades of space-age experiments. To verify some essential aspects of Einstein's theory of gravitation, NASA had commendably planned costly experiments like Gravity Probe-B, LIGO, STEP, LISA, etc. It can be naturally expected that NASA would now expeditiously plan and execute the low cost experiment proposed here from a solar probe currently under development, or from an orbiter or lander around Venus or Mercury, so as to conclusively verify the effect of the sun's gravitational field on the speed of light, as regards the important predictions of Einstein's theory of gravitation and of its remodeled form – the Remodeled Relativity Theory, which retained and incorporated only experimentally proven concepts and principles.

*PACS(2008):* 04.80.Cc, 04.50.Kd, 06.30.Gv, 03.30.+p

[*] E-mail: godopy@vsnl.com.

1. **Introduction:**

In the eighteenth and early nineteenth centuries, Newton's theory was based on three absolutes**:** space, time and mass. In the second half of nineteenth century, Maxwell made the velocity of light, c, central to electromagnetism. At the beginning of the next century, Einstein sensed that the tide of discoveries in electromagnetism indicated the decline of the mechanical view. His special relativity theory (SRT) replaced Newton's three absolutes by a single one, the velocity of light, and established two paired quantities, space-time and mass-energy, related through c. Even in the general relativity theory (GRT), which is the theory of relativistic gravitation, the velocity of light continued to play its important role. Einstein considered the speed of light derived mostly from terrestrial experiments, to be the limiting speed of all natural phenomena. But, even in the post-Einsteinian era and the space age, experimental physicists and research organizations may perhaps be said to have not focused on the issue of determining the velocity of light at locations closer to massive celestial bodies like the sun; this issue has a great importance from the point of view of the GRT as well as the Remodeled Relativity Theory (RRT) as discussed below and presented in our earlier papers.

2. **Historical Background:**

Comprehension of various aspects of light has a long history of formation of initial intuitive notions that reigned for sufficiently long periods of time, and that finally got replaced with the improved concepts after sustained efforts by theoretical and experimental physicists.

Against the current wisdom on infinite nature of the speed of light espoused by Descartes and Cassini, the Danish astronomer Ole Römer carried out the first real measurement of the speed of light. He announced to the Paris Académie des Sciences in September 1676 that the anomalous behavior of the eclipse times of Jupiter's inner moon (named Io), could be explained based on the concept of a finite speed of light. But, only after another quarter century, the scientific community was ready to accept the notion that the speed of light is not infinite, as can be seen from Newton's acceptance of the concept of finite speed of light in his book "Opticks" (1704).

The hypothesis of infinite speed of light was finally considered discredited, after an independent confirmation from observations by James Bradley (1728), who discovered the "aberration of light" and deduced that starlight falling on the Earth should appear to come from a slight angle, which could be calculated by comparing the speed of the Earth in its orbit to the speed of light. Bradley calculated the speed of light as about 298,000 kilometers per second, which is close to the currently accepted value.

Thus, the age-old intuition guided notion of physicist-philosophers on the aspect of 'infinite nature of the velocity of light' got revised and accepted by the scientific community only after half a century, and of course, the credit goes to the experimental physicists.

In 1849, Fizeau carried out the first successful measurement of the speed of light using an earthbound apparatus, employing a rapidly rotating toothed wheel. In 1862, Foucault improved on Fizeau's method by replacing the cogwheel with a rotating mirror, and estimated the speed as 298,000 km/s. During 1878-82, by improving on Foucault's method, Michelson measured very accurately the speed as 299,853 km/s, which was close to Newcomb's value.

After his triumph in measuring the speed of light so accurately, the next challenge Michelson set himself on was to detect the ether wind. But, his repeated efforts with more improved experimental setup, failed to detect any ether wind. The famous Michelson-Morley experiment conducted very accurately in 1887, gave a null result. This gave a serious blow to the prevalent ether theory, which was another age-old intuition guided notion of physicist-philosophers concerning the 'propagating medium for light' based on analogy with that for sound, and which was ultimately discarded after Einstein's presentation of SRT and its successful verifications by experimental physicists.

Almost four decades after his previous measurement of the velocity of light, Michelson in 1926, used a rotating prism to precisely measure the time taken by light to make a round trip from Mount Wilson to Mount San Antonio in California, that yielded a result of 299,796 km/s. During World War II, the development of the cavity resonance wavemeter for use in radar, together with precision timing methods, opened the way to laboratory-based measurements of the speed of light. Thus, by 1950, repeated measurements using Cavity Resonator, by Essen established a result of 299,792.5±1 km/s, which became the value adopted by the 12th General Assembly of the Radio-Scientific Union in 1957.

However, no precise local (that is, within a small region of space, where the gravitational field strength is uniform) measurement of the magnitude of "Speed of Light" at locations closer to the sun, where the solar gravitation may significantly affect it beyond the measurement uncertainties, has been proposed or reported in the relevant literature till date.

## 3. GRT and the Velocity of Light

In 1905, Einstein abandoned the ether theory and presented his SRT, which was based on the postulate of constancy of light velocity in all inertial frames. The second postulate of SRT was adopted as an extension of the Galilean relativity principle. Einstein [1] stated that it is impossible to reconcile the first postulate with the classical transformation unless it is replaced with the Lorentz transformation that is rigorously derived using the argument that if the velocity of light is same in all inertial frames, then moving rods must change their length and moving clocks must change their rhythm. However, though Lorentz transformations has served its historic role for calculating the relativistic transformation factors, it has also led to the paradoxical concept of length contraction along only the direction of motion with no contraction along the transverse direction. Even based on data from a century of relativistic experiments, there had been no direct experimental proof for the direction-dependent nature of length contraction.

Einstein later generalized the relativity theory by stating [1] that relativistic effects were caused by all kinds of energy, and that this conclusion, being quite general in character, was an important achievement of the theory of relativity and fitted all facts upon which it has been tested. But, the concept of direction-dependent length contraction leads to a paradox, as it cannot explain nature's criterion for the direction of the length contraction, for those cases where this relativistic effect is caused by any kind of energy other than the translational kinetic energy.

After formulation of the GRT in 1916, Einstein [2] had emphasized that the law of the constancy of light velocity in vacuum, which constituted one of the two fundamental postulates of the SRT, was not valid according to the GRT, as light rays could curve only when the velocity of propagation of light varies with position. He also remarked [1] that the SRT could not claim an unlimited domain of validity; as its results held only so long as it was possible to disregard the influences of gravitational fields on the phenomenon of light propagation. However, experimental verification of this important prediction of Einstein's GRT regarding the influence of gravitational fields on the velocity of light is long overdue.

### 4. Redefinition of Meter and adoption of a Fixed Value for Velocity of Light

In a bid to have a more precise 'redefinition of the meter', it was proposed to define the meter in terms of the second and a fixed numerical value, 299,792.458 m/s, for c, the speed of light [3]. Based on the prevalent concepts on the various aspects of velocity of light, the Comité Consultatif pour la Définition du Mètre (CCDM), in June of 1973, adopted the above value of c in its Recommendation M2 on the basis of measurements made up to that time. Use of this value was then recommended in Resolution 2 of the 15th Conférence Général des Poids et Mesures (CGPM) in 1975 [4]. In June of 1979 the CCDM, in Recommendation M2, used this value in proposing a new definition of the meter, based on the second:

> "The meter is the length equal to the distance traveled in a time interval of 1/299 792.458 of a second by plane Electromagnetic waves in vacuum"

Ultimately, keeping unchanged the above recommendation of the 15th CGPM, the speed of light was adopted as an official constant in 1983, at that fixed numerical value, 299,792.458 m/s, by the 17th CGPM [3]. The last experimental value was reported as 299,792,458.6 ± 0.3 m/s, based on a 1983 laser measurement by NBS, USA [5].

### 5. RRT and the Velocity of Light

RRT is an alternative relativistic gravitational model, whose theoretical and mathematical parts have been dealt in our earlier papers [6, 7, 8, 9], and whose salient features are being described below.

RRT is a remodeled form of Einstein's relativity theories, which retains and incorporates only experimentally proven principles. RRT follows the methodology of nature, and is based on a generalized law for spinning and rotational motions, which is in fact the conservation law of momentum vector direction, and which has been successfully used for the precision computation of planetary and lunar orbits. The conservation laws of energy and momentum are the most fundamental principles of RRT. Based on data from eight decades of relativity experiments, two fundamental principles were adopted for RRT: one, that energy level is the underlying cause for relativistic effects and two, that mass is expressed by Einstein's relativistic energy equation.

Utilizing the space age ephemeris generation experience and following nature's way to conserve energy and momentum, the authors found the reason to replace the concept of "relativity of all frames" with that of "nature's preferred frame", as explained in our earlier paper [6]. This helped the authors to escape the dilemma, faced by Einstein till 1912, when he concluded that there is 'no way of escape from the consequence of non-Euclidean geometry, if all frames are permissible'.

Though Einstein formulated the GRT as a law for all coordinate systems, physicists and astronomers have continued to adopt one or other 'ad-hoc' approaches that are not in complete conformity with the basic principles of Einstein's theories, as explained in our earlier paper [6].

Utilizing the essential essence of Einstein's theory, and enriched by the benefits of numerous space age experiments, RRT by its capability and consistency, can be said to have avoided the inadequacies of the former, as explained in our earlier papers [6, 7, 8, 9].

In the RRT, the magnitude of the speed of light $c_r$ varies with the radial distance from the center of the relevant gravitating mass, that is, from the origin of the "nature's preferred frame". However, the concept of variable speed of light was not adopted as any postulate of the RRT. The photon in the photon model of RRT is treated as a particle similar to any matter particle moving under the gravitational influence of the sun and the planets of the solar system, the only difference being its total energy as given by Planck's quanta, is related to its relativistic mass by Einstein's relativistic energy equation, $E = m c_r^2$.

In fact, in RRT, the light velocity has been computed using its Photon model [7], which uses an Einsteinian concept [1] that a beam of light will bend in a gravitational field 'exactly' as a material body would if thrown horizontally with a velocity equal to that of light, and thus treats the motion of photon under the influence of gravitational field at par with any material particle having mass. Also, it does not use any additional postulate or principle, but uses only those principles or basic equations that have been used by A.H. Compton in 1923, for derivations of the Compton effect. In fact, without using GRT equations this photon model rigorously proves another principle stated by Einstein [2] that the gravitational bending of light rays is the effect of variation with position of the velocity of propagation of light under the influence of gravitational field.

However, there exists one remarkable point of difference between GRT and RRT on this issue. In GRT, Einstein considered the speed of light to be the limiting speed of all natural phenomena. Thus, according to GRT, as a photon moves closer to the sun, its speed falls from the adopted standard magnitude of the speed of light. Whereas, the rigorously determined magnitude of the velocity of light at the same location, using the photon model of RRT, is higher than the standard magnitude, as obtained from a numerical simulation [7] that simultaneously computes the results for the Shapiro time delay and light deflection experiments, at their recent accuracy levels.

## 6. Discussion on Important Aspects of Einstein's relativity theory vis-à-vis the Velocity of Light

Based on the most commonly held notions by physicists and astronomers on the different aspects of velocity of light during the seventies, the numerical value of c (as mentioned at section 4 above) was fixed after a decade-long process of adoption,
- which perhaps was somewhat premature as it lacked the benefit of the new comprehensions on relativity theories that was generated by
    - the commissioning and operation of the Global Positioning System (GPS), and
    - further improvements in the space age observational as well as computational capability and accuracy in celestial mechanics,
- which was greatly influenced by Einstein's relativity theories and a few related notions.

A direct or indirect consequence of such adoption of the fixed numerical value of c, was perhaps that experimental physicists and research organizations formed the notion that space-laboratory-based measurements of velocity of light need no further experimental effort, and perhaps thus even no proposal has been made over the last quarter century, for verifying the prediction of Einstein's GRT regarding the influence of solar gravitational field (as proposed here) on the velocity of light, which surely merits to be regarded as an important test for GRT.

Some of the notions on which Einstein's relativity theories were based and several historical situations or facts that has affected the above referred commonly held notions regarding velocity of light are being briefly reviewed here. These have considerably contributed (as briefly explained in the following subsections) to formulation of the RRT and hence have been dealt in more detail in our earlier paper [6] on RRT.

### 6.1. The Concept of "Relativity of all Frames"

Einstein [1] stated that he wanted to formulate physical laws that are valid for all frames, not only for those moving uniformly, but also for those moving rather arbitrarily, relative to each other, so that it would be possible to apply such laws of nature to any frame, and that the two sentences, "the sun is at rest and the earth moves," or "the sun moves and the earth is at rest", would simply mean two different conventions concerning two different frames He added that really relativistic physics must apply to all frames and, therefore, also to the special case of the inertial frame, and hence GRT, the new general laws valid for all frames must, in the special case of the inertial system, reduce to the old laws known as the SRT. But, because of different reasons explained at the following subsections, there are two broad schools of thought among the relativists – that is, to use either GRT or only SRT, since, SRT is simpler and has been more numerously and well verified experimentally than GRT.

#### 6.1.1. Complexity of GRT mathematics

Even the two approximate methods that are used without making standard symmetry assumptions, to solve the Einstein field equations: namely, the post-newtonian approximation (p.n.a) and weak-field approximation methods – involve complex mathematics to analyze the problems. Almost a century after formulation of GRT, Stanford University in its web page [10] in connection with the Gravity Probe – B experiment, describes the crisis in the GRT in the following words: "General relativity is hard to reconcile with the rest of physics, and even within its own structure has weaknesses. Einstein himself was dissatisfied, and spent many years trying to broaden his theory and unify it with just one other branch of physics, electromagnetism. Modern physicists seeking wider unification meet worse perplexities. Above all, essential areas of general relativity have never been checked experimentally. …… Moreover, deep theoretical problems – some old and some new – remain. Einstein himself remarked that the left-hand side of his field equation (describing the curvature of space-time) was granite, but that the right-hand side (connecting space-time to matter) was sand."

### 6.1.2. Physicists using SRT in lieu of GRT

Although Einstein formulated GRT to supercede SRT, the trend of using SRT in lieu of GRT, started since the nineteen-twenties, with the relativistic electron theory of P.A.M. Dirac, relativistic quantum field theory such as QED, etc., and this trend continues to date among many physicists –

- more particularly the quantum physicists: This is because of the fact that the mathematical structures of general relativity and quantum mechanics, the two great theoretical achievements of the twentieth century physics, seem utterly incompatible [10].
- the macroscopic clock experimenters: Both Hafele-Keating [11] and Vessot et al. [12] have clearly confirmed [6] the RRT contentions that
    - time dilation occurs only for the moving clocks, which are at a higher energy level than the clocks that are at rest in the ECEF (or, the earth-centered earth-fixed), frame.
    - their data could be fitted only for the ECSF (earth-centered-space-fixed) frame, which is the "nature's preferred frame" for their experimental setup, according to the RRT.
    - their data could be fitted not by using GRT equation in the ECEF frame, but by using SRT equations in the ECSF frame (the "nature's preferred frame") while simply adding to it the gravitational effects as a separate term.
- the GPS relativists: They have also clearly confirmed the RRT contentions as experienced by the macroscopic clock experimenters mentioned above; but because of the special nature and complexity of their problem, these are being presented in little more detail in the next subsection.

### 6.1.3. Problems faced by GPS relativists

The GPS relativists faced wider perplexities while commissioning their state-of-the-art timekeeping system that may be considered a continuous or ongoing clock experiment over a wider region of space. As the ECEF frame is of primary interest for navigation and the GPS operates in the gravitational field of the earth, GRT is thus supposed to be the relativity model that is appropriate to tackle the gravitational problem in a rotating ECEF frame. Thus, Ashby mentioned [13] that an early GPS design decision was taken to broadcast the satellite ephemerides in a model ECEF frame, designated by the symbol WGS-84. He further mentioned [13] that it became ultimately necessary to invoke the inertial frame (the ECSF frame, that is, the "nature's preferred frame" according to the RRT) for GPS application, for the following reasons:

- the receiver needed to generally perform a different rotation for each measurement made, into some common inertial frame, so that the navigation equations apply, and after solving the propagation delay equations, a final rotation must usually be performed into the ECEF to determine the receiver's position. He concluded that this became exceedingly complicated and confusing, needing them to switch over to the ECSF frame [13].
- Simple-minded use of Einstein synchronization in the rotating frame leads to a significant error, because of the additional travel time required by light to catch up to the moving reference point, as compared to the travel time required in the underlying inertial frame [13].
- many physical processes (such as electromagnetic wave propagation) are simpler to describe in an inertial frame [13].

- the inertial frames are certainly needed to express the navigation equations (that is, the four simultaneous one-way signal propagation equations), as it would also lead to serious error in asserting these equations in the rotating ECEF frame [13].
- in a rotating reference frame, the Sagnac effect prevents a network of self-consistently synchronized clocks from being established by transmission of electromagnetic signals [14].
- as a consequence of the Sagnac effect, observers in the rotating ECEF frame using Einstein synchronization will not agree that clocks in the ECSF frame are synchronized, due to the relative motion [14].
- observers in the rotating frame cannot in fact even globally synchronize their own clocks, due to the rotation of the earth [14].

**6.1.4. Physicists using only GRT**

The astrophysicists, astronomers and some physicists are falling in this group.

### 6.1.4.1. The physicist/ astronomers

The astronomers working on ephemeris generation are using GRT though not in its purest form, and find it useful to employ GRT equations in a "preferred frame" of the gravitating mass, for planetary and lunar orbits. They find that the LSA derived astronomical constants (e.g., the planetary masses, etc.) of nature, which are an outcome of global fits done during the generation of a particular ephemeris, are consequences not only of the gravitational model, but also of the coordinate frame. In other words, the constants of nature are linked to the coordinate frame, which means that today one has to accept the existence of the constants of nature as a concomitant of only one appropriate preferred frame (which has been termed as the "nature's preferred frame", according to the RRT) and the relevant orbit or orbits, linked to them.

### 6.1.4.2. Astrophysicists

The astrophysicists use GRT but not in its purest form, in the sense that they also need to resort to a "preferred frame" of the gravitating mass.

RRT has replaced the concept of "relativity of all frames" with that of the "nature's preferred frame", and employed the same to consistently and successfully simulate numerically the results of all the "well-established" tests of the GRT at their current accuracy levels [7, 8, 9], and for the precise calculation of relativistic effects observed in case of the GPS applications, the accurate macroscopic clock experiments and other tests of the SRT [6].

**6.2. Lorentz transformations**

This is one of the relativistic notions that relates to and strongly influences the commonly held notions regarding velocity of light. A few related important aspects may be briefly reviewed here.
- Lorentz transformation was developed from electromagnetism, and derived based on rectilinear motion in an inertial frame, from a thought experiment involving propagation of the electromagnetic wave itself. The concept of the constancy of the velocity of light is embedded in its very derivation method, as is clear from Einstein's question [1] in this connection – "Can we not assume such changes in the rhythm of the moving clock and in the length of the moving rod that the constancy of the velocity of light will follow directly from these assumptions?" To which he himself replied by reversing the same argument, and stating that if the velocity of light is the same in all frames, then moving rods must change their length, moving clocks must change their rhythm, and the laws (*Lorentz transformation laws*) governing these changes are rigorously determined [1].
- Einstein [1] also stated that it is impossible to reconcile the first postulate of his special relativity theory (SRT) with the classical transformation unless it is replaced with the Lorentz transformation, and that the number expressing the velocity of light appears explicitly in the Lorentz transformation, and plays the role of a limiting case, like the infinite velocity in classical mechanics.

- Thus, the role of Lorentz transformation in relativity theory got over-emphasized, because of the following reasons**:**
    - it provided the earliest method for the theoretical calculation of the transformation factor for the relativistic effects like the kinematical time dilation, length contraction, etc., and
    - it had to play the important role for eliminating the contradiction between the first and the second postulates of SRT.
- Based on the data from a century of relativistic experiments, it has not been possible to answer the question**:** what is nature's criterion for the direction of the length contraction, for those cases where the relativistic effect is caused by any kind of energy other than the translational kinetic energy.
- No direct experimental test has confirmed the paradoxical direction-dependent nature of the phenomenon of length contraction over almost a century.

RRT has discarded this paradoxical transformation law and replaced it with the energy level based transformation factor based on the experimentally proven principle that energy level (due to any form of energy) is the more fundamental underlying cause behind the relativistic effects [6].

### 6.3. The principle of constancy of the speed of light in vacuum

Since, SRT is simpler and can be learnt more intuitively, by the time the would-be-physicists pass to the learning stage of GRT, the principle of constancy of the speed of light in vacuum and similar other intuitive principles become deeply embedded in the mind of even many renowned physicists, as an important and inescapable part of their relativistic concepts. C.M. Will mentions [15] that during his lecture in 1961, Irwin Shapiro (discoverer of the famous Shapiro Time delay, the fourth test of GRT) was puzzled on hearing that speed of light is not constant in GRT, when Shapiro, five years after his Ph.D. in physics from Harvard university was engaged at MIT's Lincoln laboratory, for improved determination of the astronomical unit by radar ranging to planets, and had already concluded that radar ranging could be used to give improved measurements of Mercury's perihelion shift. Will further mentions [15] that Shapiro was puzzled because he had always thought that according to relativity the speed of light should be the same in every inertial frame, though he knew that GRT predicts that light should be deflected by a gravitating body, but the question that remained unanswered in his mind was**:** Would its speed also be affected. Will also mentions [15] how Shapiro applied his corrected notion on the speed of light for estimating the round-trip time of a radar signal to a distant object, and the result was the discovery of the Shapiro Time delay effect. This is an example of a real life situation when eliminating an incorrect or unproven notion related to the concepts on the speed of light or the relativity theory, can be useful for improving ones understanding and making new discoveries.

In fact, this approach to eliminate unproven (or, incorrect) notions with proven (or, correct) concepts has helped the authors considerably in the remodeling effort for RRT, which consist of only experimentally proven concepts. And, based on a similar approach, in this paper the authors are proposing that a long-held notion about an unexplored aspect of the velocity of light, be placed on the anvil of experimental test that will ultimately deepen every relativist's comprehension of relativity theory.

### 6.4. Velocity of light is the upper limit of velocities for all material bodies

At the beginning of twentieth century, when Einstein was framing up his relativity theory, the line of thinking amongst his contemporary physicists was expressed by Poincaré [16], who delivered in 1904 a lecture titled "The Principles of Mathematical Physics," at the International Congress of Arts and Sciences in St. Louis, during which he mentioned in connection with relativistic concepts**:** "Perhaps, likewise, we should construct a whole new mechanics of which we only succeed in catching a glimpse, where inertia increasing with the velocity, the velocity of light would become an impassable limit". At that epoch, physics was ready to switch from the concept of 'absolute time', as is evident from a 1904 paper (presenting his transformation law) by H.A. Lorentz where he introduced a novel term called "local time",
- which was used by him to explain the motion of matter through ether, and
- which according to him, was a kind of mathematical trick for simplifying the Maxwell equations, and was having no experimental meaning [16].

Einstein, who introduced the path-breaking concept of relativistic time, and who used the same transformation law, eliminated the need of such "mathematical trick" of Lorentz. Such a historic role was played by the Lorentz transformation law and its link with the concept of the velocity of light being the upper limit of velocities for all material bodies, was thus fixed from the necessity of giving birth to the SRT. Thus, Einstein himself made the statement [1] that the number expressing the velocity of light appears explicitly in the Lorentz transformation, and plays the role of a limiting case, like the infinite velocity in classical mechanics. Another great spin-off from the SRT is Einstein's relativistic energy equation, $E= mc^2$, which has been quantitatively verified many times over the last century, but, all such experiments had been conducted at terrestrial laboratories, where the numerical value of c correspond to 299,792.458 m/s. And, one must take note of the fact that the strength of earth's gravitational field is so low that if a photon travels up to the limit of earth's sphere of influence, its velocity varies from the fixed numerical value by only 0.2 m/s, according to both GRT and RRT.

However, during the formulation of SRT and GRT, it was widely recognized that the velocity of light forms the upper limit of velocities for all material bodies, and thus Einstein made the statement [1] that the velocity of light formed the upper limit of velocities for all material bodies, and that the simple mechanical law of adding and subtracting velocities is no longer valid or, more precisely, is only approximately valid for small velocities, but not for those near the velocity of light. This apparently simplistic statement needs an in depth review after a century of advancement of experimental physics, which has taught us that the conservation laws of energy, and of linear and angular momentum are the more general forms of the laws of mechanics, and are more fundamental than the laws of mechanics in the sense that they are valid in both the macroscopic and microscopic realms. Einstein made the above statement, because the simple mechanical law of adding and subtracting velocities assumes that mass as constant, whereas at velocities near the velocity of light, the relativistic mass becomes enormously high. Next, Einstein stated [1] that the number expressing the velocity of light appears explicitly in the Lorentz transformation, and plays the role of a limiting case. Thus, it is clear that the Lorentz transformation law dictated the upper limit of velocities for all material bodies.

It may however, be mentioned here that since RRT adopted Einstein's relativistic energy equation, $E= mc_r^2$, as one of its fundamental principles, $c_r$ forms the upper limit of velocities for all material bodies at coordinate r in the relevant "nature's preferred frame". And, in RRT, the velocity for any material body is rigorously determined by applying the conservation laws of energy and momentum [6]. In contrast, the GRT notion that the numerical value of c, measured from terrestrial experiments forms the upper limit of velocities for all material bodies anywhere in the universe, can be logical only if the value of c measured at locations closer to massive bodies like the sun, as in case of the experiment proposed here, is found to be lower than the fixed numerical value of c.

**6.5. Space-age Solution to Einstein dilemma before accepting non-Euclidean geometry**

As elaborated in our earlier paper [6], Einstein faced a dilemma till 1912 considering the consequence of bringing in non-linear transformations for generalizing the Lorentz transformations to include accelerations [16], but during the pre-space age; he had no way out but to conclude that 'there is no way of escape from the consequence of non-Euclidean geometry, if all frames are permissible'.

Whereas, while formulating RRT, the authors got the advantage of experimental results of space age, and could find the reason to replace the concept of "relativity of all frames" with that of "nature's preferred frame", which helped us
- to escape the dilemma, faced by Einstein till 1912,
- to find out the deeper underlying cause of relativistic effects, and develop the energy-level based transformation law,
- to avoid use of the Lorentz transformation law, which led to the experimentally unproven and paradoxical direction-dependent nature of the phenomenon of length contraction, and
- to avoid the non-Euclidean geometry.

### 6.6. RRT Methodology and an Important Aspects of Velocity of Light

#### 6.6.1. Salient Points on RRT Methodology

In RRT, the authors
- did not include any postulate or experimentally unverified principle,
- included only experimentally proven principles,
- used in its Photon model an Einsteinian concept [1] that a beam of light will bend in a gravitational field 'exactly' as a material body would if thrown horizontally with a velocity equal to that of light, and thus treats the motion of photon under the influence of gravitational field at par with any material particle having mass,
- took advantage of improvements in technological and computational capabilities, followed the way nature is applying the conservation laws in case of a real phenomenon of nature, and utilized in RRT this methodology of nature, by incorporating a continuous check for conformance to the conservation laws of energy, and of linear and angular momentum. This approach helps us to identify not only the "nature's preferred frame", but also the real phenomenon from an apparent phenomenon.
- used Euclidean space to consistently and successfully simulate numerically the results of all the "well-established" tests of the GRT at their current accuracy levels (as presented in our earlier papers), and for the precise calculation of relativistic effects observed in case of the GPS applications, the accurate macroscopic clock experiments and other tests of the SRT, and
- found that the precisely computed value of the velocity of light proves another principle stated by Einstein [2] that the gravitational bending of light rays is the effect of variation with position of the velocity of propagation of light under the influence of gravitational field.

#### 6.6.2. GRT versus RRT computed values of the velocity of light

The RRT computed value of the velocity of light differs from that obtained using GRT. In GRT, Einstein considered the speed of light to be the limiting speed of all natural phenomena. Thus, according to GRT, as a photon moves closer to the sun, its speed falls from the numerical value of c fixed at 299,792.458 m/s which was adopted by the 17th CGPM in 1983, and which corresponds to a value determined at terrestrial laboratories. As mentioned above,
- Einstein's derivation of the Lorentz transformation started with the SRT postulate of the constancy of the speed of light, and
- Einstein stated [1] that the number expressing the velocity of light appears explicitly in the Lorentz transformation, and plays the role of a limiting case.

These starting assumptions or postulates has led to the situation that GRT-computed value of c falls below the upper limit, as a photon moves closer to the sun.
Whereas, the rigorously determined magnitude of the velocity of light at the same location, using the photon model of RRT, is higher than the standard magnitude, as obtained from a numerical simulation [7] that simultaneously computes the results for the Shapiro time delay and light deflection experiments, at their recent accuracy levels.
However, only an experiment as proposed in this paper can prove once for all whether the RRT contention mentioned above is true or not.

### 7. Discussion on the Proposed Experimental Test for the Speed of Light

Since the earliest determinations of speed of light, many experiments have been conducted till accuracy reached the level of a fraction of a meter per second. But, by 1983, the speed of light was defined to be a constant of nature, and its magnitude was adopted as 299,792,458 meters per second. It was too early a period in the space age. However, the result of the proposed experimental test would not necessitate any change in the already adopted fixed value of c, or in the definition of the meter adopted by the 17th CGPM in 1983, even if such result proves the RRT value given below. Because, even for RRT, the reference value of c has been adopted as 299,792,458 m/sec. at mean sea level (MSL) on earth [6].
In view of the discussions above, the authors propose an experiment that should verify the magnitude of the velocity of light at locations closer to the sun at an accuracy level of at least fractional metre per second.

Such an experiment can be expeditiously planned and executed from an orbiter around or lander (depending on requirements from the point of view of convenience of experimental set up, and cost considerations) on any one of the planets**:** Mercury or Venus. According to GRT, the measured value will be lower than the standard magnitude by about 7 metres per second for the case of Mercury, and by about 4 metres per second for the case of Venus. Alternatively, such an experiment can be expeditiously incorporated in the mission objective for the solar probe mission currently under development under NASA's plans.

According to the RRT, the measured value will be higher than the standard magnitude by about 7 metres per second for the case of Mercury, and by about 4 metres per second for the case of Venus.

Thus, the experimental result will clearly prove whether GRT or RRT is the right relativistic gravitational model.

NASA had commendably planned and executed a very costly and long-term experiment like Gravity Probe-B, to verify a hitherto unverified (by any direct experiment) but essential aspect of Einstein's theory of gravitation. NASA had spent excellent effort on direct experimental detection of gravitational radiation, another hitherto unverified (by any direct experiment) and essential aspect of Einstein's GRT, using LIGO. NASA had in the planning and developmental stage, a few other programs like STEP, LISA, Solar Probes etc. It can be naturally expected that NASA or similar other research organizations would now plan and execute the experiment proposed here that will verify from this low cost experiment, another hitherto unverified and essential aspect of Einstein's GRT

### 8. Conclusion

In view of the above, it is expected that NASA or similar other research organizations would now come forward and actively plan for the proposed experiment, in view of the facts that this test
- will ultimately deepen every relativist's comprehension of relativity theory and relativistic effects,
- will provide an experimental verification of an important prediction of Einstein's GRT regarding the influence of gravitational fields on the velocity of light, which is overdue for long,
- will verify whether the GRT notion that the numerical value of c, measured from terrestrial experiment forms the upper limit of velocity for any material body located anywhere in the universe, and
- will be one of the lowest cost space-age test for GRT as compared to the ones presently under execution and planning, viz., like Gravity Probe-B, LIGO, STEP, LISA, Solar Probes etc.